\documentclass[12pt]{spieman}  
\usepackage{amsmath,amsfonts,amssymb}
\usepackage{graphicx}
\usepackage{setspace}
\usepackage{tocloft}
\usepackage{pifont}
\newcommand{\cmark}{\ding{51}}%
\newcommand{\xmark}{\ding{55}}%
\usepackage{lineno}
\title{Switching Loss for Generalized Nucleus Detection in Histopathology}
\author[a,*]{Deepak Anand}
\author[b]{Gaurav Patel}
\author[a]{Yaman Dang}
\author[a,c]{Amit Sethi}
\affil[a]{Electrical Engineering, Indian Institute of Technology Bombay, Mumbai MH 400076, India}
\affil[b]{Electronics and Communication Engineering, National Institute of Technology Raipur, Raipur CH 492010, India}
\affil[c]{Department of Pathology, University of Illinois at Chicago, Chicago IL 60612, USA}

\cftpagenumbersoff{figure}
\cftpagenumbersoff{table} 
\begin{document} 
\maketitle
\begin{abstract}
The accuracy of deep learning methods for two foundational tasks in medical image analysis -- detection and segmentation -- can suffer from class imbalance. We propose a `switching loss' function that adaptively shifts the emphasis between foreground and background classes. While the existing loss functions to address this problem were motivated by the classification task, the switching loss is based on Dice loss, which is better suited for segmentation and detection. Furthermore, to get the most out the training samples, we adapt the loss with each mini-batch, unlike previous proposals that adapt once for the entire training set. A nucleus detector trained using the proposed loss function on a source dataset outperformed those trained using cross entropy, Dice, or focal losses. Remarkably, without retraining on target datasets, our pre-trained nucleus detector also outperformed existing nucleus detectors that were trained on at least some of the images from the target datasets. To establish a broad utility of the proposed loss, we also confirmed that it led to more accurate ventricle segmentation in MRI as compared to the other loss functions. Our GPU-enabled pre-trained nucleus detection software is also ready to process whole slide images right out-of-the-box and is usably fast.
\end{abstract}


{\noindent \footnotesize{*}Address all correspondence to Deepak Anand,  \linkable{deepakanand@iitb.ac.in} }

\begin{spacing}{1}   
\section{Introduction}
\label{intro}

Detection of each nucleus on a slide as a separate entity is an important task in computational pathology because pathologists routinely base their diagnosis on the size, shape, and spatial distribution of stain density in individual nuclei and the spatial relationships between nuclei~\cite{elston_pathological_1991,stierer_nuclear_1992,dunne_scoring_2001}. For example, hyperplasia is diagnosed based on a high nuclear packing density in a tissue slide stained with the most commonly used stain -- hematoxylin and eosin (H\&E). Detection of individual nuclei also plays an important role in the use of many immunohistochemistry (IHC) panels and in fluorescence microscopy, where cells with specific markers need to be counted. Furthermore, nucleus detection also makes segmentation of individual nuclei for morphometric studies much easier~\cite{pinidiyaarachchi2005seeded,kowal2019cell}.

Methods that give satisfactory nucleus detection accuracy right out-of-the-box, that is, without retraining or fine-tuning, on target datasets can lower the barrier to using computational methods in digital pathology. However, the appearance of nuclei and their surroundings can vary so significantly by organs, scales, datasets, patients, and stains, that nucleus detection without fine-tuning on a target dataset has been attempted only once before~\cite{Naylor_distancemap2019}. Even that method was tested on a single target dataset that shared the organ, magnification, and stain with the source dataset that was used for training.

We propose a method that detects nuclei more accurately without retraining on target datasets than previous methods even though those methods were trained, at least partially, on the same target datasets or organs. In our experiments the target datasets not only represent separate hospitals, but also separate organs, and stains than those represented in the source dataset used for training our method.

Although object segmentation and detection are well-studied for natural images ~\cite{girshick2014rich,girshick2015fast,ren2015faster,redmon2016you,he2017mask}, instance-level nucleus detection in histopathology is challenging due the following reasons:
\begin{itemize}
\item \emph{Class imbalance:} The ratio of foreground to background pixels is usually severely tilted towards the latter, especially for nucleus detection, where only the pixels that are close to the centroids of nuclei are considered foreground. Additionally, the foreground to background ratio varies from image to image depending on the grade of the disease and the sampling location in a tissue. By contrast, usually the object to be detected in a natural image is quite salient.
\item \emph{Large number of instances with varied appearances:} Each histopathology image can have hundreds nuclei whose appearance varies across images by organs, cell-types, disease states, patients, hospitals, reagent compositions, scanners, magnifications, and lab protocols. The number of instances of objects to be detected in natural images is usually much smaller.
\item \emph{Crowding and occlusion:} Nuclei tend to crowd together in many disease conditions, such as hyperplasia and high-grade cancers, which makes it difficult for algorithms to detect individual nuclei as separate instances. Similarly, in natural images, pedestrian counting in crowded urban scenes is still considered difficult~\cite{Wan_2019_CVPR}.
\end{itemize}

The novelty of our method lies in solving the problems of instance-level nucleus detection mentioned above. Firstly, we propose a new loss function to address the problem of class imbalance. Although class imbalance in semantic segmentation has been tackled via weighted loss functions in the recent past, these weighing schemes depend on the ratio of foreground to background pixels in the entire training set, which is fixed across training mini-batches~\cite{khan2017cost,lin2018focal}. Such a weighing scheme leads to a sub-optimal training of a model that may perform well on the held-out images of one dataset but fail to accurately detect nuclei without retraining on another dataset. We propose a novel `switching loss' function that not only adaptively shifts the emphasis between foreground and background, but is also resets for each training mini-batch to make better use of the training data for better generalization. We also tested that the proposed loss function is applicable to problems other than nucleus detection, and indeed it improves performance of models to segment right ventricle in MRI images compared to the other losses. Secondly, to solve the problem of generalizing over varied shapes and appearances of nuclei, we used a large and diverse source dataset for generalization across organs and hospitals~\cite{kumar2017convolutional,kumar_dataset_2017}. Additionally, to generalize to staining variations and new stains, we used stain separation as a pre-processing step~\cite{vahadane_structure-preserving_2016,sethi2016empirical}. Furthermore, to train model that are robust to changes in the size of the nuclei, we extensively used scale augmentation. And lastly, we rigorously compared several neural network architectures for instance-level nucleus detection.
 
We solved two additional practical challenges to make our method readily usable by the computational pathology community. Firstly, digital scanning of glass slides with an equivalent 40x objective lens, which correspond to 0.25 $\mu m$ of tissue distance per pixel dimension, can yield upward of 3.5 Giga (x$10^9$) pixels per whole slide image (WSI). Secondly, most of the slide scanners output files in proprietary multi-resolution TIFF formats. Available image processing libraries can neither handle very large images, nor open images in these formats. We wrote an open-source software utility that applies our nucleus detection method directly on WSI files by using the OpenSlide library~\cite{goode2013openslide}. The code of the software utility is also GPU-optimized such that it seamlessly detects all nuclei in each gigapixel WSIs within six minutes using the widely available Nvidia Titan X GPU. That is, ours is the first open-source software that can directly analyze WSIs, let alone operating reasonably fast.

The rest of the paper is organized as follows. We review the existing works and literature in Section~\ref{sec:bck}. In Section~\ref{sec:theory}, we present the overall approach. In Section~\ref{sec:results}, we present the comparative analysis of the proposed method with state-of-the-art methods and conclude the article in Section~\ref{sec:discuss}.

\section{Background and Previous Work}
\label{sec:bck}
In this section, we will review previous work on nucleus detection, fully convolutional networks, and the problem of class imbalance in medical image segmentation.

\subsection{Previous work on nucleus detection}

Early attempts at the nucleus detection have been mainly based on hand-crafted features that capture the intensity and texture variations in the images, such as Otsu thresholding~\cite{phansalkar_adaptive_2011}. Later, binary morphological filtering using structural elements such as a circle, square, star, etc. have been used to detect nuclear shapes~\cite{plissiti_automated_2011}. Clustering to find super-pixels and region growing to refine nuclear boundaries for segmentation have also been proposed~\cite{sarrafzadeh_nucleus_2015}. Another approach used for nucleus segmentation is to use active contour methods, which are deformable splines that represent the boundary of an image segment using gradient information~\cite{sadr_leukocytes_2010}. Graph-cut for partitioning the graph of pixel intensities into sub-graphs, where the similarity within the subgraphs has also been tried~\cite{qi_drosophila_2013}. The settings used in these methods are very dependent of the target dataset, which limits the breadth of their applicability, especially to situations where nuclei are crowded, enlarged, or varied in appearance.

With the advent of deep learning and datasets with thousands of annotated nuclei, nucleus detection accuracy has improved in recent years. However, the challenges such as crowded nuclei, staining variations, and variations due to disease state remain. Innovations to separate touching nuclei include predicting the boundary and interior of the nuclei as separate classes~\cite{kumar2017convolutional,yuxin_countour_2018,graham2018xy,hofener2018deep,Yanning_CIA}, applying watershed algorithm on predicted nuclear probability map~\cite{naylor2017nuclei}, and applying shape deformation methods~\cite{xing2015automatic}. Predicting the distance transform with respect to the centroid of the nucleus is another approach~\cite{distance_transform,kumar2017convolutional}. One may also use the boundary refinement block within a deep learning framework~\cite{peng_large_2017} to solve this issue. These methods still require post-processing that is both computationally costly and has settings that are dependent on dataset. To speed up the inference on WSIs, we aimed to reduce post processing to just non-maximal suppression of the predicted output map. And, to speed up training for detection, we adopt a very efficient method for ground truth preparation that involves shrinking the individual nuclei masks.

We did not compare with approaches based on generative adversarial networks (GANs) because we do not find them to be practical for large multi-organ, multi-scale datasets. These approaches are very specific to the distribution of the training data and often don't generalize well due to the difficulty of finding a Nash equilibrium for the min-max game being played. Instead of introducing artificial variations in the training data that may or may not be biologically plausible, we seek diversity from a large multi-organ dataset with scale augmentation.

\subsection{Fully convolutional networks for detection and segmentation}

\begin{figure}
    \centering
    \includegraphics[width=12cm]{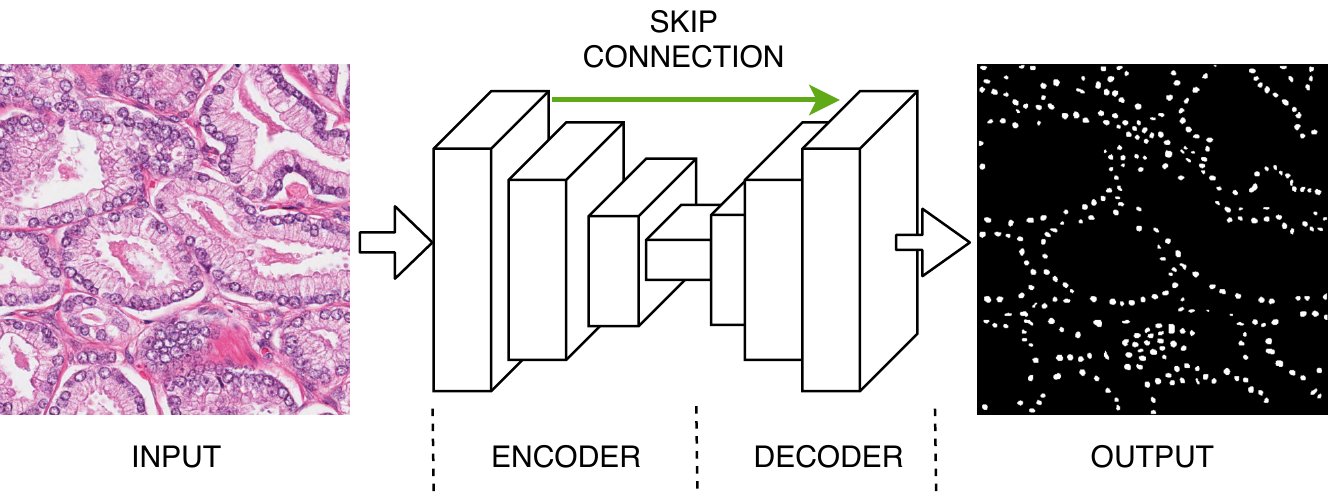}
    \caption{A general representation of fully convolutional networks: The encoder is composed of convolutional and pooling layers for downsampling and the decoder is composed of deconvolutional layers for upsampling. Best viewed on a color monitor.}
    \label{fig:fcn}
\end{figure}

With advances in the architectures of convolutional neural networks, it is now clear that fully convolutional network offer much better processing speed over the patch-based approaches~\cite{long2015fully}. Due to the absence of fully connected layers, these models can take images of different sizes as input and generate a segmentation map (e.g., foreground vs. background label for each pixel).  Although there are various architectures for FCNs, most models can be generically represented as shown in Figure \ref{fig:fcn} using two basic blocks -- one for encoding (downsampling) and another for decoding (upsampling). The encoder is responsible for extracting features at a large spatial scale (e.g., those pertaining to large objects) from the input using learned convolutional filters. Max-pooling layers are used to increase the spatial context between convolutional layers. Its output is a feature map of smaller spatial size. The decoder block is responsible for constructing the class probability map with the same spatial resolution as that of the input sub-image. It uses upsampling layers using transposed 2D convolution. Additionally, for finer context, it often uses skip connections to read the features computed by the layers of the encoder at the matching resolution ~\cite{ronneberger_u-net:_2015}. The last layer of the decoder generates the output class probability map.

\subsection{Class imbalance in medical image segmentation}

Most of the segmentation and detection problems in medical images suffer from class imbalance, which has been primarily addressed by data sampling strategies~\cite{lee2016plankton,pouyanfar2018dynamic,buda2018systematic} and loss functions that try to restore the balance in favor of the under-represented class~\cite{khan2017cost,lin2018focal}. Data sampling strategies leave large portions of the training data unused thus we take route of modifying loss. Some popular loss functions for segmentation and detection are binary cross entropy (BCE) and Dice loss. The BCE loss $L_C$ expression is shown in equation \ref{eq:bce}
\begin{equation}
\label{eq:bce}
    L_{C} = -\sum_{i} \sum_{j} \log{p(i,j)}
\end{equation}
where, $p(i,j)$ is the predicted probability that the pixel at location $(i,j)$ belongs to the ground truth mask of a nucleus. The BCE loss tries to match the background and foreground pixels in the prediction to the ground truth masks, which results in erroneous salt and pepper prediction pattern, i.e., foreground pixels in the background region or holes in the intended foreground region. Moreover, BCE does not address class imbalance.

The Dice loss $L_D$ is an alternative to BCE that considers the entire object by computing the overlap between predicted and ground truth objects, and is described in equation \ref{eq:dice}
\begin{equation}
\label{eq:dice}
    L_{D}=1- \frac{2\sum_{i} \sum_{j} p(i,j)\  g(i,j) + \epsilon}{\sum_{i} \sum_{j} p(i,j) + \sum_{i} \sum_{j} g(i,j) + \epsilon}
\end{equation}
where, $g(i,j)$ is the corresponding binary ground truth label, and $\epsilon$ is a small constant to prevent division by zero. The Dice loss considers only the foreground pixels while calculating the overlap between the prediction and the ground truth. Due to the variation in nucleus packing density across organs and disease states, the inter-nuclear separation varies widely between images. Using only the Dice loss for regions with crowded nuclei incurs only a small penalty for false-positives. On the other hand, if the Dice score is computed only for the background -- when it is called inverted Dice loss $L_I$ -- then it has the opposite disadvantage in case of well-separated and small nuclei. Inverted Dice loss is described as follows.
\begin{equation}
    L_{I}=1- \frac{2\sum_{i} \sum_{j}(1- p(i,j))\ (1-g(i,j)) + \epsilon}{\sum_{i} \sum_{j} (1-p(i,j)) + \sum_{i} \sum_{j} (1-g(i,j)) + \epsilon}
\end{equation}

A common modification in the loss function is to increase the weight of the loss corresponding to the less abundant foreground class. On similar lines, focal loss $L_F$ was proposed where the loss assigned to well-classified pixels is reduced by using a multiplicative term with the cross entropy loss. Focal loss is shown in equation \ref{eq:focal}
\begin{equation}
\label{eq:focal}
    L_{F} = -(1-p(i,j))^{\gamma}\log{p(i,j)}
\end{equation}
where, $\gamma$ is a hyper-parameter to control the emphasis on the misclassified pixels. 

The previously proposed loss modification techniques to counter class-imbalance suffer from two major issues. Firstly, they are very focused on the foreground objects, i.e., weighing for the background is ignored. Secondly, the weight for the foreground class is set once for the entire training sample, and batch-wise variations are ignored. We address these two problems while proposing the `switching loss' function.

\section{Proposed Method}
\label{sec:theory}

\begin{figure*}[!h]
    \centering
    \includegraphics[trim={0 13.3cm 0 0},clip,width=14cm]{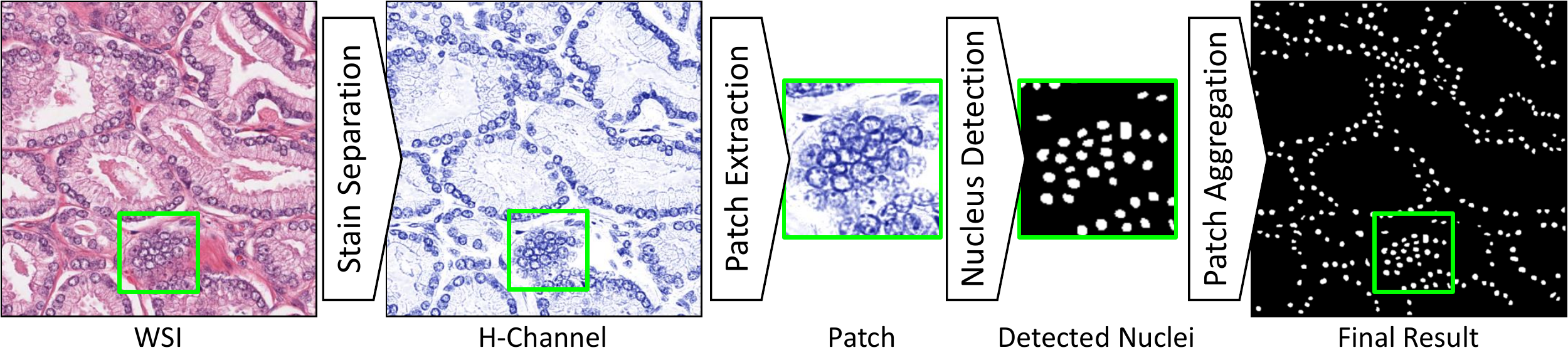}
    \caption{Flowchart of nucleus detection: The stains in the input WSI are separated to obtain the target nuclear stain density (e.g. hematoxylin). Patches are extracted from the stain density map. Nucleus center probability map is computed by an FCN. Resultant patches are stitched together to yield a WSI level map followed by non-maximal suppression for the exact nuclear locations. Best viewed on a color monitor.}    
    \label{fig:flowchart}
\end{figure*}

We first give an overview of our method before going into the details of each step. As shown in Figure \ref{fig:flowchart}, the proposed method is based on end-to-end fully convolutional neural networks (FCNs) and involves minimal post-processing to speed up inference. To improve generalization to new datasets, we input stain density map -- obtained by stain separation -- to the neural network instead of the raw image. Ground truth masks of individual nuclei for training images were the central 25\% area of the nuclei, which was often equivalent to a single pixel at 20x and a few pixels at 40x magnification. To adapt to the resultant class imbalance due to different nuclear densities, we propose to train the FCN using the proposed `switching loss' function that shifts the emphasis given to false positives and false negatives depending on the nuclear packing density. We now describe the rationale behind these design choices in more detail.

\subsection{Weakly annotated ground truth mask}

\begin{figure}[!h]
    \centering
    \includegraphics[width=4cm]{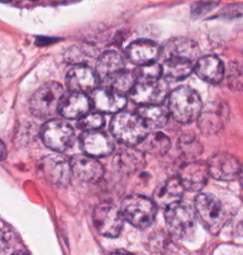}
    \includegraphics[width=4cm]{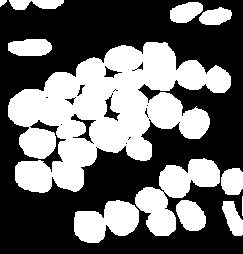}
    \includegraphics[width=4cm]{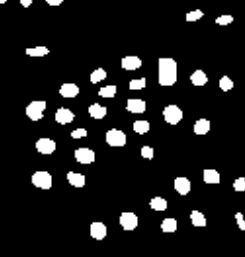}
    \caption{Ground truth generation: In images with crowded nuclei (left) masks of individual nuclei overlap and merge (middle), but shrinking the masks radially can lead to separated ground truth masks for individual nuclei even when binarized (right). Best viewed on a color monitor.}
    \label{fig:shrinking}
\end{figure}

We adopt an efficient approach during training to mitigate the problem of crowded nuclei and overlapping nuclei. We use the central 25\% are of an annotated nucleus as the ground truth foreground region. This technique gave well-separated ground-truth masks that aided detection of individual nuclei, as shown in Figure \ref{fig:shrinking}. Segmentation is an easy extension of detection as the detected centroids can be used as seeds for a segmentation step~\cite{pinidiyaarachchi2005seeded}.

\subsection{Switching loss function}

We counter the issues associated with class imbalance by proposing a new loss function for training nucleus detectors, which is a sum of BCE loss $L_C$ and an adaptive combination of Dice loss $L_D$ and inverted Dice loss $L_I$. We switch of the emphasis between Dice and inverted Dice depending on the ratio of the number of nuclear pixels $C_n$ and the total number of pixels $C_t$ in a mini-batch as compared a threshold ($\tau$). If the ratio is lower than the threshold then more importance is given to true positives in comparison to the true negatives, and vice versa. The proposed loss function can be described using the following expression:

\begin{equation}
L = \begin{cases}
L_{C} +\lambda \  L_{D} + (1-\lambda)\  L_{I},& \text{ for } \frac{C_n}{C_t}>\tau \\
L_{C} +(1-\lambda) \  L_{D} + \lambda\  L_{I},& \text{ for } \frac{C_n}{C_t}<\tau
\end{cases}
\label{eqn:loss_new}
\end{equation}
where, the hyperparameters have the following ranges:
\begin{equation*}
 {0 \leq \lambda \leq 1} \text{ and }
 0 \leq \tau \leq 1
\end{equation*}
The weighing by the hyperparameter $\lambda$ or $1-\lambda$ gave higher emphasis to the under-represented class between the nuclei and the background.

\subsection{GPU-enabled fast stain separation}

\begin{figure}[t]
    \centering
    \includegraphics[width=6cm]{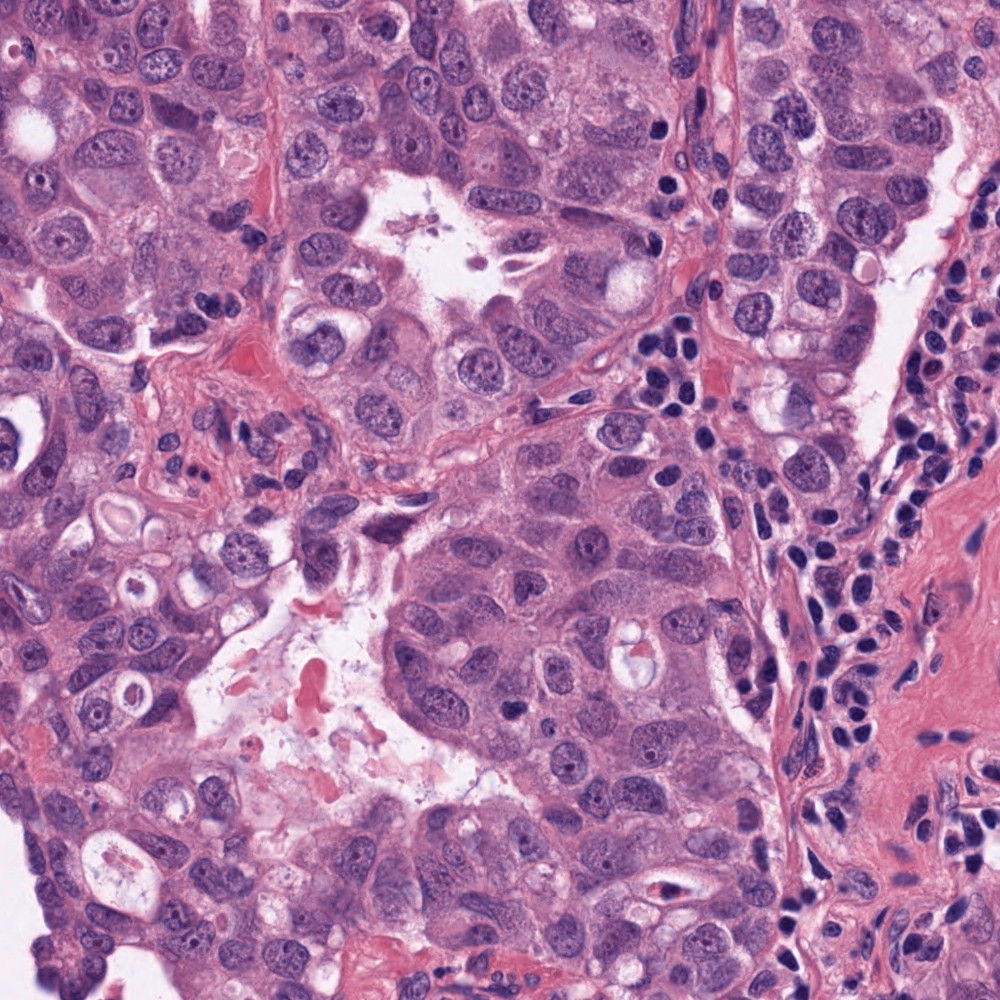}
    \includegraphics[width=6cm]{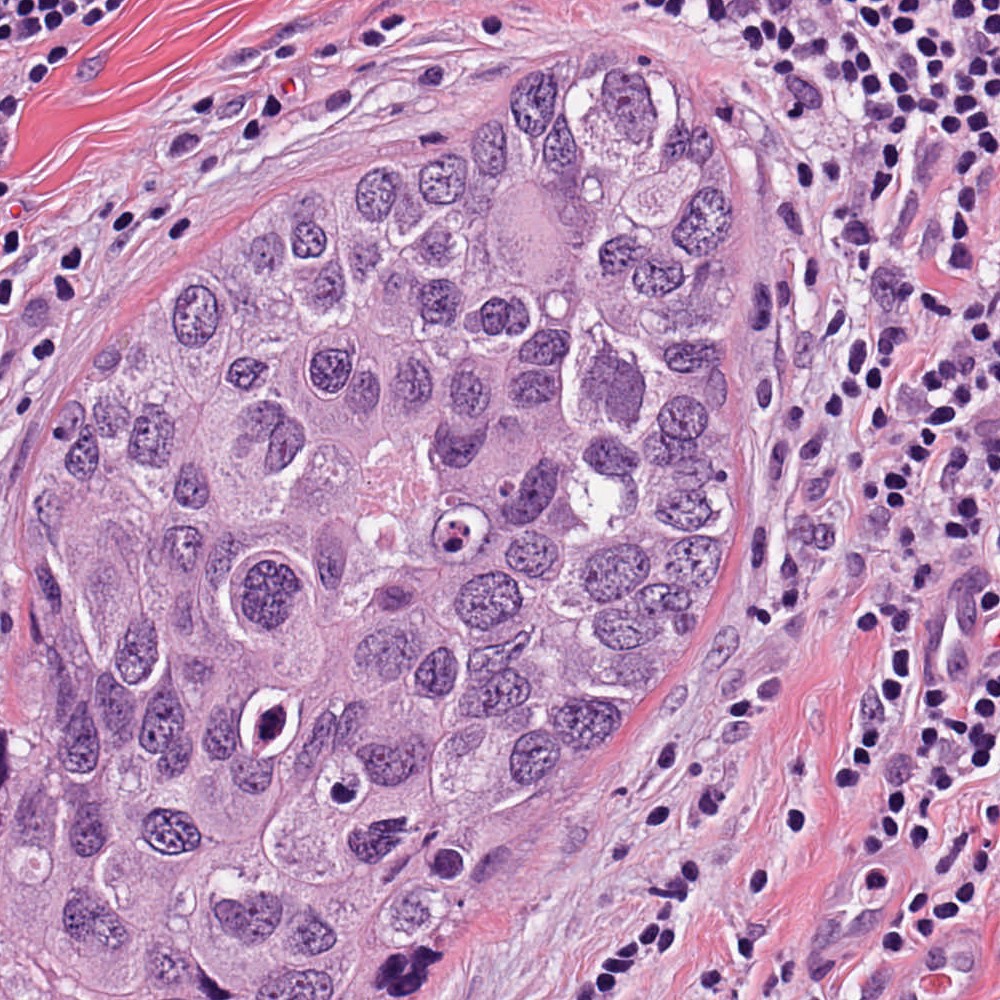}
    \caption{Sample images of lung (left) and breast (right) cancer tissues from the multi-organ dataset (MOD)~\cite{kumar_dataset_2017} show how nuclei across organs can vary in appearance. Best viewed on a color monitor.}
    \label{fig:sample_mod}
\end{figure}

One of the major challenges in nucleus detection is the variation in color even with H\&E staining due to variation in factors such as reagent compositions and scanners sensor response, as evident from the images in Figure \ref{fig:sample_mod}. It has already been established that using raw (unnormalized) digital pathology images lowers pixel classification accuracy~\cite{sethi2016empirical}. We considered the choice between using color normalized images or the nuclear stain (H-channel) images as input to the neural network, and found the latter to be more advantageous. Nuclear stain images not only led to more accurate results in our experiments, training a neural network on these images made it generalize to unseen stains (e.g. immunohistochemistry) without retraining.

We use an efficient implementation of structure-preserving color normalization (SPCN)~\cite{vahadane_structure-preserving_2016} that was extended for GPUs~\cite{ramakrishnan_fast_2019}. It uses sparse non-negative matrix factorization for stain separation and is compatible with OpenSlide library~\cite{goode2013openslide} that we wanted to use to read proprietary WSI file formats. SPCN is an unsupervised method for stain-transfer and is well-suited for our purpose due to its proven validity of stain density estimation and computational efficiency.

\begin{figure}[!h]
    \centering
    \includegraphics[width=4.2cm]{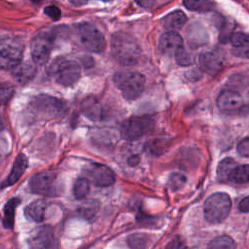}
    \includegraphics[width=4.2cm]{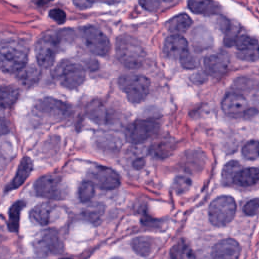}
    \includegraphics[width=4.2cm]{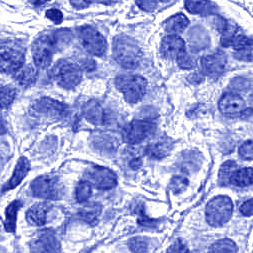}
    \caption{Importance of color-normalisation step. Nucleus-dense epithelium regions (left) when color normalized (middle) may still not have the desired visual contrast between nuclei and their surroundings, whereas viewing only the H-channel of the stain-separated image (right) increases the visual contrast. Best viewed on a color monitor.}
    \label{fig:sample_clornorm}
\end{figure}

While the color-normalization step is helpful for nucleus detection, it cannot resolve the problems with dense epithelial regions with no pinkish intervening stroma, as can be seen in Figure \ref{fig:sample_clornorm}. In such regions, we found that using H-stain images works slightly better than color-normalization, as has also been observed previously~\cite{ahmad_correlation_2018}. Stain density estimates are inherent intermediate outputs of SPCN.

\subsection{Fully convolutional networks}

To make the inference fast we decided to use FCN architectures for nucleus detection. We experimented with the following four FCN architectures that are popular for image segmentation and compared their results:

\begin{itemize}

\item SegNet: This architecture is one of the first neural network models explicitly designed for semantic segmentation~\cite{badrinarayanan_segnet:_2015}. To perform upsampling in the decoder, it uses $\text{argmax}$ coordinates for max-pooling layers of the encoder so that there are no learnable parameters needed for upsampling. While this saves on the number of learnable parameters, our experiments show that it leads to slightly lower accuracy.
    
\item UNet: This architecture is one of the most widely used ones for semantic segmentation~\cite{ronneberger_u-net:_2015}. It uses skip connection between encoder and decoder blocks, concatenating (appending at the end channel-wise) the feature map from the encoder block to the decoder block feature map. Skip connections facilitate context-aware construction of the higher resolution output and a smoother flow of the gradients to the earlier layers (without dilution) during backpropagation to produce sharp segmentation maps.

\item VGG-UNet: A UNet architecture with a VGG11 encoder is called VGG-UNet~\cite{iglovikov_ternausnet:_2018}. A VGG11 encoder that is pre-trained on ImageNet dataset is often used. The learned features from the encoder block are concatenated to the decoder block. The weights of encoder block are fine-tuned instead of learning from scratch. This results in faster convergence of the training process.

\item Global Convolution Networks (GCN): GCN uses large convolutional kernels instead of stacking many small ($3\times3$) kernels to gather larger spatial context for segmentation tasks ~\cite{peng_large_2017}. It also uses a boundary-refinement block based on a residual unit at the end, which gives refined boundaries at the edges of the objects.
    
\end{itemize}

Apart from using FCNs we ensured that we used minimal post-processing to improve the inference speed even further. We simply compare the predicted nucleus center probability map to a threshold of 0.35 and compute the centroid for the predicted contour to detect the nuclei. The threshold value was empirically determined once on the validation set, and then set for all the other experiments.

\section{Experiments and Results}
\label{sec:results}
In this section, we evaluate the performance of our approach on two publicly available datasets with annotated nuclei to demonstrate its state-of-the-art performance. We also show several qualitative results, including those on stains, organs, and datasets that our trained neural network had not seen before. To demonstrate the general applicability of the proposed loss and training method, we also achieved state-of-the-art results for segmenting right ventricle in an MRI dataset.


\subsection{Nucleus detection on Multi-Organ Dataset (MOD)}

The first dataset used in our experiments is a multi-organ dataset that consists of 30 H\&E stained sections taken from WSIs different patients covering seven different organs at 40x magnification~\cite{kumar_dataset_2017}. Each section is 1000$\times$1000 pixels in size. The annotations for individual nuclear boundaries have been provided. The composition for the MOD dataset is shown in Table \ref{tab:desc_mod}. We prepared a training dataset (TD) and a validation dataset (VD) separated by patients from the slides of four organs -- breast, prostate, lung, and kidney. For testing, we prepared two datasets. The first consisted of only organs used for training (but a different set of patients) and is referred to as TeD1. The other test dataset referred to as TeD2, comprised of images from unseen-organs \& different patients to test a stronger generalization. The details of the split for the formation of these datasets can also be found in Table \ref{tab:desc_mod}. A few sample images taken from this dataset are shown in Figure \ref{fig:sample_mod}.
\begin{table}
\caption{Details of the multi-organ dataset (MOD)~\cite{kumar_dataset_2017}: MOD was split into training (TD), validation (VD), test 1 (TeD1), and test 2 (TeD2) subsets that were separated by patients. TeD2 is further separated by organs as well.}
\label{tab:desc_mod}
\begin{center}
\begin{tabular}{|c|c|c|c|c|c|c|c|}
    \hline
    \# Images $\downarrow$, Organ $\rightarrow$ & Prostate & Kidney & Breast & Lung & Bladder & Colorectal & Stomach \\
    \hline
    Training Dataset (TD) & 3 & 3 & 3 & 3 & - & - & -\\
    Validation Dataset (VD) & 1 & 1 & 1 & 1 & - & - & -\\
    Testing Dataset 1 (TeD1) & 2 & 2 & 2 & 2 & - & - & -\\
    Testing Dataset 2 (TeD2) & -& - & - & - & 2 & 2 & 2\\
    \hline
    Count & 6 & 6 & 6 & 6 & 2 & 2 & 2\\
    \hline
\end{tabular}
\end{center}
\end{table}

\begin{table}[!h]
\caption{F1-score comparing our results to previous works using the conditions used in the respective papers.}
\label{tab:40x}
\begin{center}
\begin{tabular}{|c|c|c|}
    \hline
    Test  &  Previous Work & F1 score \\
    dataset &  & \\\hline
      &  VGG-UNet with Switching loss (Equation \ref{eqn:loss_new}) & 0.854 \\
      & Yuxin et.al. (2018) ~\cite{yuxin_countour_2018} & 0.850\\
      &  VGG-UNet with BCE + Dice + Inverted Dice & 0.843 \\
      &  VGG-UNet with BCE + Dice & 0.839 \\
    MOD 40x  & Yanning et.al. (2019) ~\cite{Yanning_CIA} & 0.837\\
    (TeD1 +TeD2)   &  VGG-UNet with Focal Loss  & 0.833 \\  
      & Zeng et.al (2019) ~\cite{zeng_ric_unet} & 0.828\\
      & Kumar et.al.(2017) ~\cite{kumar_dataset_2017} & 0.827\\
      & Naylor et.al (2019) ~\cite{Naylor_distancemap2019} & 0.787\\
      & Qu et.al. (2019) ~ \cite{qu2018weakly} & 0.778\\
      & Graham et.al. (2018) ~\cite{graham2018xy} & 0.773\\
    \hline 
\end{tabular}
\end{center}
\end{table}

\subsection{Training on MOD dataset}
To account for staining variations across datasets, we passed only the H-channel images to the FCN instead of the original images. To separate touching nuclei, we shrunk their ground truth masks to 25\% area as mentioned previously. After this step, we augmented the images and their masks using random rotations and flips. To train our method for multi-scale nucleus detection, the training images were augmented by scaling down by up to a factor of $4\times4$. Since these images were too large to be fed into the neural network, we extracted overlapping patches of size $256\times256$ from the augmented images and their corresponding ground truth masks. We coded the networks in PyTorch~\cite{paszke2017automatic} and trained them for 100 epochs with a batch size of 4 and the initial learning rate of $10^{-4}$, which decayed by a factor of 0.3 every 25 epochs. At the time of testing, our code extracts patches from WSIs on the fly in a raster fashion and feeds them to the FCNs. To save processing time, we filter out patches that lack the tissue segments by checking if each patch contains more than 60\% white pixels (intensity$>$220 on a 256 level scale). Nuclei masks are obtained from the FCN and passed to a patch aggregation block that builds a whole WSI mask from the coordinate locations in each patch. The locations of the nuclei are simply the centroids of the connected components obtained from the mask image. There is no computationally expensive post-processing required, such as erosion, dilation, or watershed segmentation, to separate touching nuclei due to the way training ground truth mask was generated by shrinking masks of individual nuclei.

We compare the performance of a trained VGG-UNet model on MOD test dataset(TeD1+TeD2) by training it  using various loss functions -- a) the proposed switched loss, b) equal weighing of Dice \& inverted Dice with BCE, c) only Dice with BCE and d) focal loss \cite{lin2018focal} with $\gamma$ equal to 5. The comparisons are shown in Table \ref{tab:40x}. We achieved state-of-the-art F1-score outperforming every other method, which were more sophisticated in terms of training and post-processing. This performance is the result of the proposed pipeline and the balanced training facilitated by the proposed switching loss. We can also see an increasing F1-score as we use more elements of the proposed loss function, thus indicating the utility of Dice, inverted Dice, and adaptive weighing of the two based on nuclear density.
 
\begin{figure}
    \centering
    \includegraphics[width=6cm]{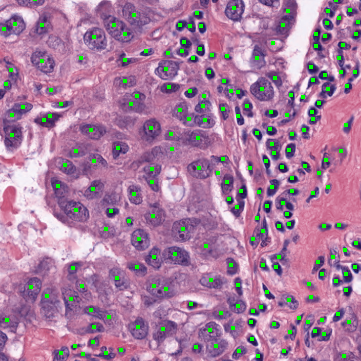}
    \includegraphics[width=6cm]{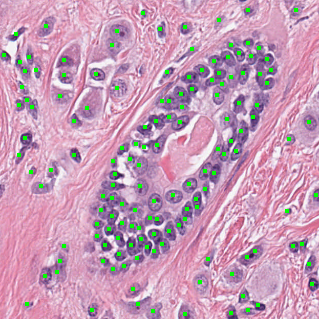}
    \caption{Visual test results on MOD~\cite{kumar_dataset_2017} (green dots represent the detected nuclei) demonstrate generalization across organs (the image one the left is of a lung, while the one on the right is that of a breast tumor). Best viewed on a color monitor.}
    \label{fig:tmi}
\end{figure}

We present visual results on the MOD dataset in Figure~\ref{fig:tmi}. The detection performance in the crowded region is noteworthy.

\subsection{Generalization on an independent dataset}
\begin{figure}[!h]
    \centering
    \includegraphics[width=6cm]{tnbc/581910_7.png}
    \includegraphics[width=6cm]{tnbc/498959_03.png}
    \includegraphics[width = 6cm]{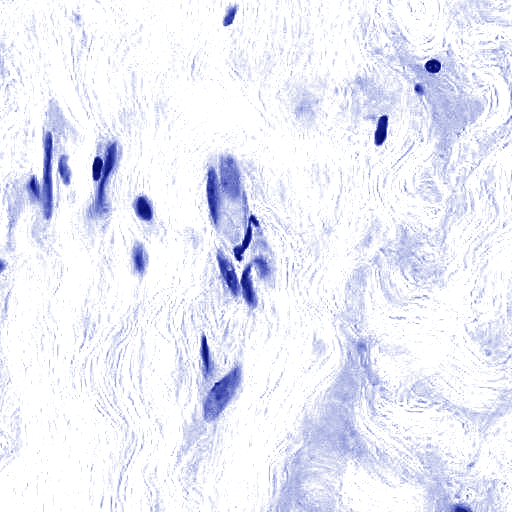}
    \includegraphics[width =6 cm]{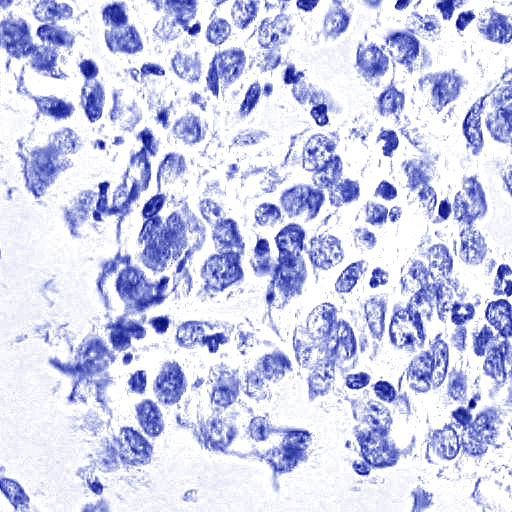}
    \caption{Sample images from TNBC datasets with their corresponding H-channel images. Note the diversity in the dataset and the apparent advantages of the stain-separation for input standardization. Best viewed on a color monitor.}
    \label{fig:tnbc}
\end{figure}

To demonstrate the generalization of the proposed framework and trained models right-out-the-box we evaluate the MOD-trained model on TNBC dataset without retraining or fine-tuning on the latter. TNBC dataset was released by ~\cite{naylor2017nuclei}. TNBC dataset contains images from triple negative breast cancer. This dataset contains 33 H\&E images of size 512 $\times$ 512 pixels containing a total of 2754 annotated cells. It contains a considerable number of annotated cells, including normal epithelial and myoepithelial breast cells, invasive carcinoma cells, fibroblasts, endothelial cells, adipocytes, macrophages, and inflammatory cells (lymphocytes and plasmocytes). Thus TNBC dataset captures a wide variety of cells. It presents a challenging and diverse dataset to work with, as evident from Figure \ref{fig:tnbc}. The results shown in Table~\label{tab:tnbc} demonstrate that VGG-Unet trained on H-channel images of MOD dataset using the proposed switching loss on TNBC dataset outperforms other networks, even though our network was not fine-tuned on the TNBC dataset while the other networks trained on a subset of the TNBC dataset images.

\begin{table}[!h]
\caption{Comparison on TNBC dataset: F1-score comparing our results to previous works using the conditions used in the respective papers. The proposed method did not train or fine-tune on any subset of images of the TNBC dataset unlike the other networks.}
\label{tab:tnbc}
\begin{center}
\begin{tabular}{|c|c|c|}
    \hline
    Test  &  Previous Work & F1 score \\
    dataset &  & \\\hline
    All images of TNBC  &  Ours  & 0.830\\
    (without training)  & Naylor et.al (2019) ~\cite{Naylor_distancemap2019} & 0.807\\
    \hline
      & Naylor et.al (2019) ~\cite{Naylor_distancemap2019} (trained on only breast images from MOD) & 0.823\\
    Test Images    & Naylor et.al (2019) ~\cite{Naylor_distancemap2019} (trained on TNBC and MOD) & 0.811\\
    of TNBC  & DeconvNet Naylor et.al (2017) ~\cite{naylor2017nuclei} (trained on TNBC) & 0.805\\
      & FCN  Naylor et.al (2017) ~\cite{naylor2017nuclei} (trained on TNBC) &  0.763\\
      & PangNet Naylor et.al (2017) ~\cite{naylor2017nuclei} (trained on TNBC) & 0.676\\
    \hline
\end{tabular}
\end{center}
\end{table}

\subsection{Qualitative results}
In this section we show qualitative results of the proposed model on other H\&E datasets and also on other (immunohistochemistry) stains to establish the scope and impact of the work outside of the datasets and stains used in the previous experiments.

\subsubsection{Performance on other H\&E datasets}
\begin{figure}
\centering
    \includegraphics[width=6cm]{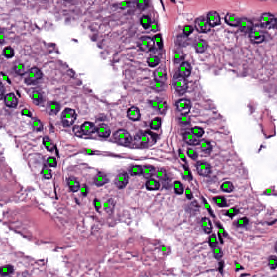}
    \includegraphics[width=6cm]{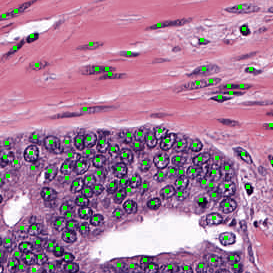}
    \caption{Visual results of testing on CPCTR dataset (green dots indicate detected nuclei) demonstrate generalization across nuclear densities. Best viewed on a color monitor.}
    \label{fig:cpctr}
\end{figure}

\begin{figure}
    \centering
    \includegraphics[width=6cm]{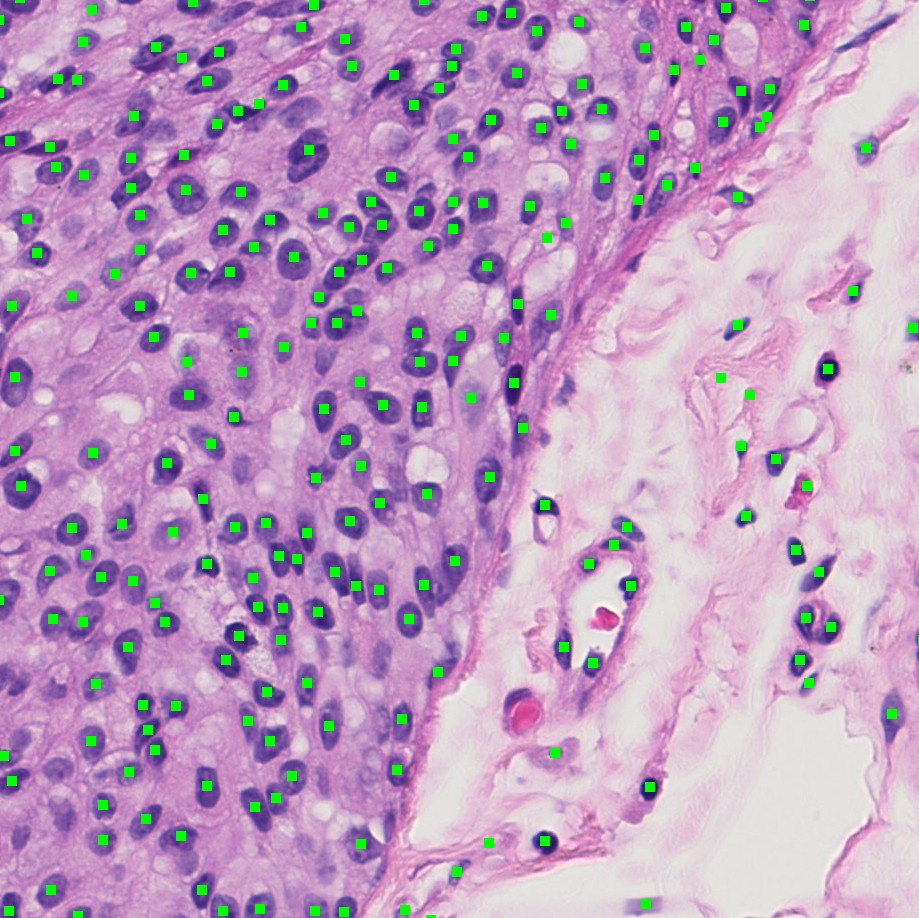}
    \includegraphics[width=6cm]{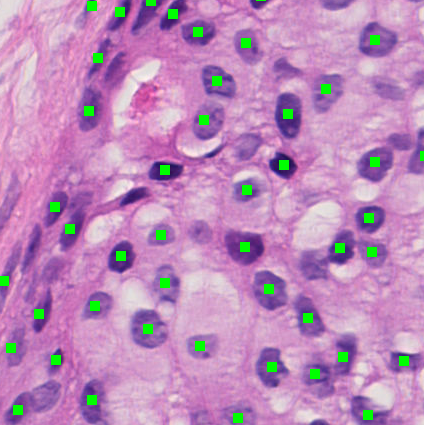}
    \caption{Visual test results on Warwick dataset (green dots indicate detected nuclei) demonstrate generalization across nuclear morphologies and magnifications (image on the left is at 20x, while the one on the right is at 40x). Best viewed on a color monitor.}
    \label{fig:warwick}
\end{figure}

\begin{figure}
    \centering
    \includegraphics[width=6cm]{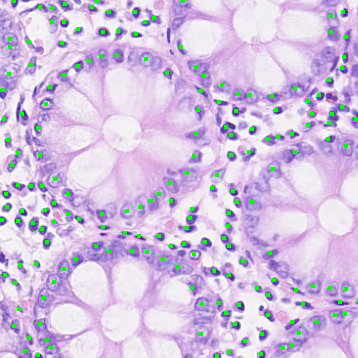}
    \includegraphics[width=6cm]{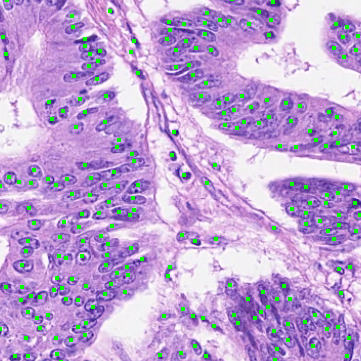}
    \caption{Visual test results on CRC dataset (green dots represent the detected nuclei) demonstrate generalization across disease states (the image on the right has a high grade tumor). Best viewed on a color monitor.}
    \label{fig:crcd}
\end{figure}

We now present visual results obtained using our method on various H\&E-stained datasets. Figure \ref{fig:crcd} shows sample results on the CRC dataset~\cite{sirinukunwattana_locality_2016} that demonstrate generalization across disease states. Generalization across organs can also be seen in the results on MOD~\cite{kumar_dataset_2017} as shown in Figure \ref{fig:tmi}. The ability of our method to detect nuclei right out-of-the-box can also be seen visually on two other datasets that do not have ground truth annotations. The first dataset was sourced from Cooperative Prostate Cancer Tissue Resource (CPCTR)~\cite{melamed_cooperative_2004}, which consists of H\&E stained prostate cancer tissue microarray scanned at 20x resolution. The second dataset was sampled from a HER2 scoring competition hosted by the University of Warwick~\cite{qaiser_her2_2018}, which consists of H\&E images from breast cancer at 40x resolution. Figure \ref{fig:cpctr} shows results on CPCTR dataset, in which our method ensures separate detection of individual nuclei even in the most crowded regions. The results on Warwick dataset in Figure \ref{fig:warwick} show generalization across nuclear morphology. We want to re-emphasize that our model detects a wide-range of nuclei, including those in the stroma, epithelium, adipose, and the lymphatic system. In several datasets, some of these nuclei, such as lymphocytes and those of the adipose tissue, are not even marked. We leave the choice to the user to filter out those types of nuclei that may not be important for their study based on morphological or learned filters.

\subsubsection{Performance on immunohistochemistry images}

\begin{figure}
    \centering
    \includegraphics[width=6cm]{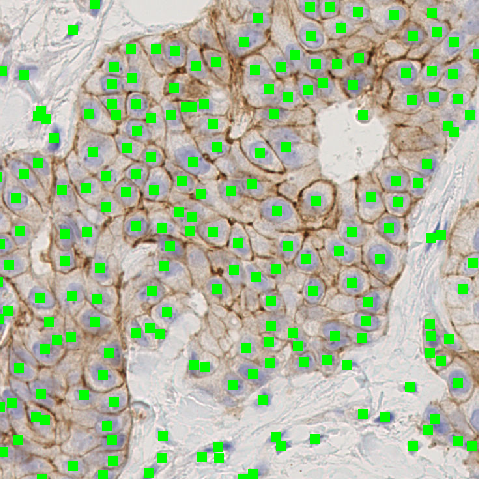}
    \includegraphics[width=6cm]{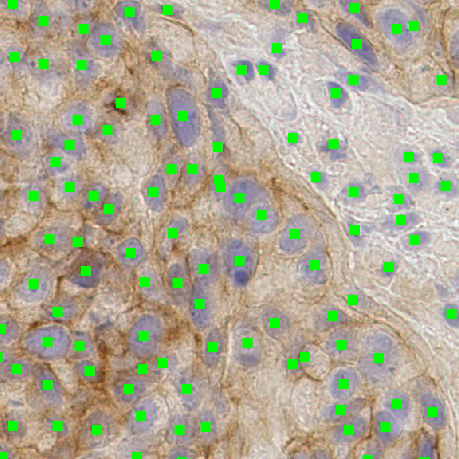}
    \includegraphics[width=6cm]{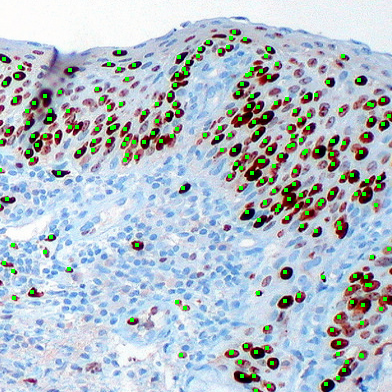}
    \includegraphics[width=6cm]{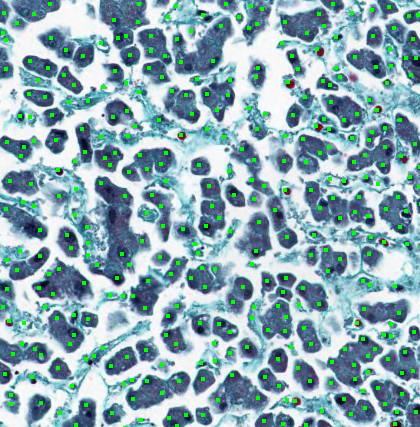}
    \caption{Visual test results on IHC slides (green dots indicate detected nuclei) demonstrate generalization to unseen stains. The two images on the top are for HER2 IHC of breast cancer in which nuclei counter-stained with hematoxylin (blue) were to be detected and the one on the left-bottom is that of Ki-67 immunostain used on cervical tissue for which positive nuclei (brown) were to be detected. The left-bottom image is a IHC stained image with extremely merged nuclei and non-contrasting stains. Best viewed on a color monitor.}
    \label{fig:ihc}
\end{figure}

Our proposed solution also works for IHC images with only minor adjustments to the code but without retraining the FCNs. For a membranous IHC stain for which nuclei are counter-stained with hematoxylin, nucleus detection is straightforward using our method. The hematoxylin (blue) and IHC (brown) channels were separated automatically using modified SPCN~\cite{vahadane_structure-preserving_2016,ramakrishnan_fast_2019} and the former was fed into the FCN. The first two sub-figures of  Figure \ref{fig:ihc} present visual results on breast cancer HER2neu IHC counter-stained with hematoxylin that was sampled from the Warwick HER2 scoring contest dataset~\cite{qaiser_her2_2018}. 

We also present results on an image where a nuclear IHC stain was used. The results in the second-last sub-figure of Figure~\ref{fig:ihc} are those for Ki-67 IHC in cervical tissue, where the positive (affected) nuclei appear brown due to their adherence to the staining antibody, and the negative ones absorb only the hematoxylin counter-stain. We show that our method can detect even the IHC-stained positive (brown) nuclei by passing the IHC-channel by exchanging the basis vectors of separated stains with the change of a single line of code. Thus, our pre-trained model can also be used for counting IHC positive nuclei even when they are not stained with hematoxylin.

We also present results on an image which has a very distinctive IHC staining (see last sub-figure \ref{fig:ihc}). This image is challenging as it poses problems for stain-separation and extreme merging of nuclei. As evident from the figure, our pre-trained model can separate merged nuclei and performs satisfactorily in detecting nuclei.

\subsection{Ablation Studies with popular FCN architectures and normalization approaches}

We trained four architectures, i.e., SegNet, UNet, VGG-UNet, and GCN on training dataset (TD) using switching loss and multi-scale augmentation. The value of $\lambda$ was set to 0.8, and $\tau$ was set to 0.2 for multi-organ dataset after empirical observations. We formed two variants of the dataset. The first variant had color-normalized H\&E images, and the second one had H-channel images. We trained a separate model for each image variant and each FCN architecture to compare their relative usefulness.

Table \ref{tab:TeD1} shows the performance on both test datasets for all models. First two rows give a summary of the F1-score achieved by different architectures on the dataset with images from the same organs as those used for training (lung, kidney, breast \& prostate), i.e., TeD1. This dataset is multi-scale (i.e., contains images on various magnification level ranging from 10x to 40x), and the performance is reported on the entire dataset. CN-H\&E stands for the strategy where the input image is color-normalized and fed into the model. H-C stands for the strategy where we hematoxylin channel obtained after stain separation as input to the trained model. VGG-UNet architecture gave the best performance for both H\&E and H images and outperformed previous results on this dataset even though those results were for a single scale (40x)~\cite{kumar_dataset_2017}.

The last two rows summarize F1-score achieved by different architectures on unseen-organs of TeD2 dataset that consists of multi-scale images from the bladder, colorectum, and stomach. GCN model performed best for both H\&E image and H-channel image input. The performance of all the models on this dataset was slightly lower than that on the seen-organ dataset (TeD1), which was expected.

\begin{table}
\caption{F1-scores comparing four FCN architectures and two pre-processing schemes -- color-normalization and H-channel separation (H-C). Results were averaged across magnifications ranging from 10x to 40x and images for two test datasets -- one with the same organs (TeD1) and another with different organs (TeD2) as those in the training set.}
\label{tab:TeD1}
\begin{center}
\begin{tabular}{|c|c|c|c|c|c|}
    \hline
    Dataset  &  Pre-  &  SegNet  &  UNet  &  VGG-  &  GCN \\
     &  processing  &   &   &  UNet  &  \\ \hline
    TeD1 (seen organ)  &  CN-H\&E & 0.831 & 0.852 &  0.855  & 0.846\\
    TeD1 (seen organ) &  H-C  & 0.850 & 0.854 &  0.857  & 0.856\\
    TeD2 (unseen organ) &  CN-H\&E & 0.798 & 0.816 & 0.812 &  0.824 \\
    TeD2 (unseen organ) &  H-C  & 0.815 & 0.831 & 0.830 &  0.836 \\ \hline
\end{tabular}
\end{center}
\end{table}

\subsection{Applicability of the switching loss to MRI segmentation}

To demonstrate the utility of the proposed switching loss for class-imbalanced semantic segmentation tasks beyond nuclei, we use switching loss to train a right ventricle segmentation model on a cardiac MRI dataset. We used MICCAI 2012 right ventricle segmentation challenge(RVSC) dataset~\cite{petitjean2015right}. The challenge was to automatically segment right ventricle endocardium and epicardium segmentation from short-axis cine MRI. The ground truth data is given in the form of contours for the endocardial and epicardial area. There are 16 patients in the training set. We split this training set in 12:4 for training and testing the segmentation model. The results mentioned below are for endocardium segmentation only.

\begin{figure}[!h]
    \centering
    \includegraphics[width=16cm]{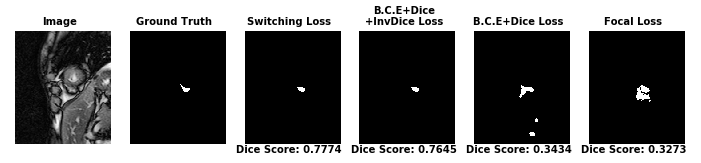}
    \includegraphics[width=16cm]{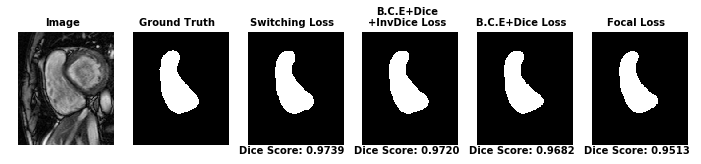}
    \caption{Sample images from RVSC datasets with their corresponding ground truth and predicted segmentation mask images. Notice the class-imbalance in the dataset and the better performance of switching loss in terms of Dice score on the image.}
    \label{fig:radiology}
\end{figure}

We train a U-Net~\cite{ronneberger_u-net:_2015} on the frames of the cine MRI using four different loss functions a) proposed switching loss b) equal weighing of Dice \& inverted Dice with BCE, c) only Dice with BCE and d) focal loss \cite{lin2018focal} with $\gamma$ equal to 1. Figure \ref{fig:radiology} shows sample images from the dataset with their corresponding ground truth and predicted mask using various loss functions. As evident from Figure \ref{fig:radiology}, the RVSC is a class imbalanced segmentation challenge and proposed loss performs better than other losses. Thus, the proposed switching loss is a good candidate for training the segmentation models.

Table \ref{tab:loss_comp_radiology} shows the average Dice score achieved on the validation dataset by using various loss functions and comparison of our results with state-of-the-art deep learning frameworks on the same datasets. The proposed loss function performs better than other loss functions and achieves state-of-the-art.

\begin{table}[!h]
\caption{Comparison of average Dice-scores achieved using different loss functions and previous methods on endocardium RVSC validation dataset.}
\label{tab:loss_comp_radiology}
\begin{center}
\begin{tabular}{|l|c|}
    \hline
    Method & Dice-Score \\ \hline
    UNet with Switching loss (Equation \ref{eqn:loss_new} with $\lambda =0.75$) & 0.863 \\
     Luo et.al (2016)~\cite{shortaxis_2nw} & 0.860 \\
    UNet with BCE + Dice + Inverted Dice & 0.853 \\
    Borodin et.al(2018)~\cite{Unet_part_dil} & 0.850\\
    UNet with BCE + Dice & 0.849 \\
    UNet with Focal Loss  & 0.824 \\
    \hline
\end{tabular}
\end{center}
\end{table}

\subsection{Optimal $\lambda$ for the proposed switching loss}

\begin{figure}[!h]
    \centering
    \includegraphics[width=12cm]{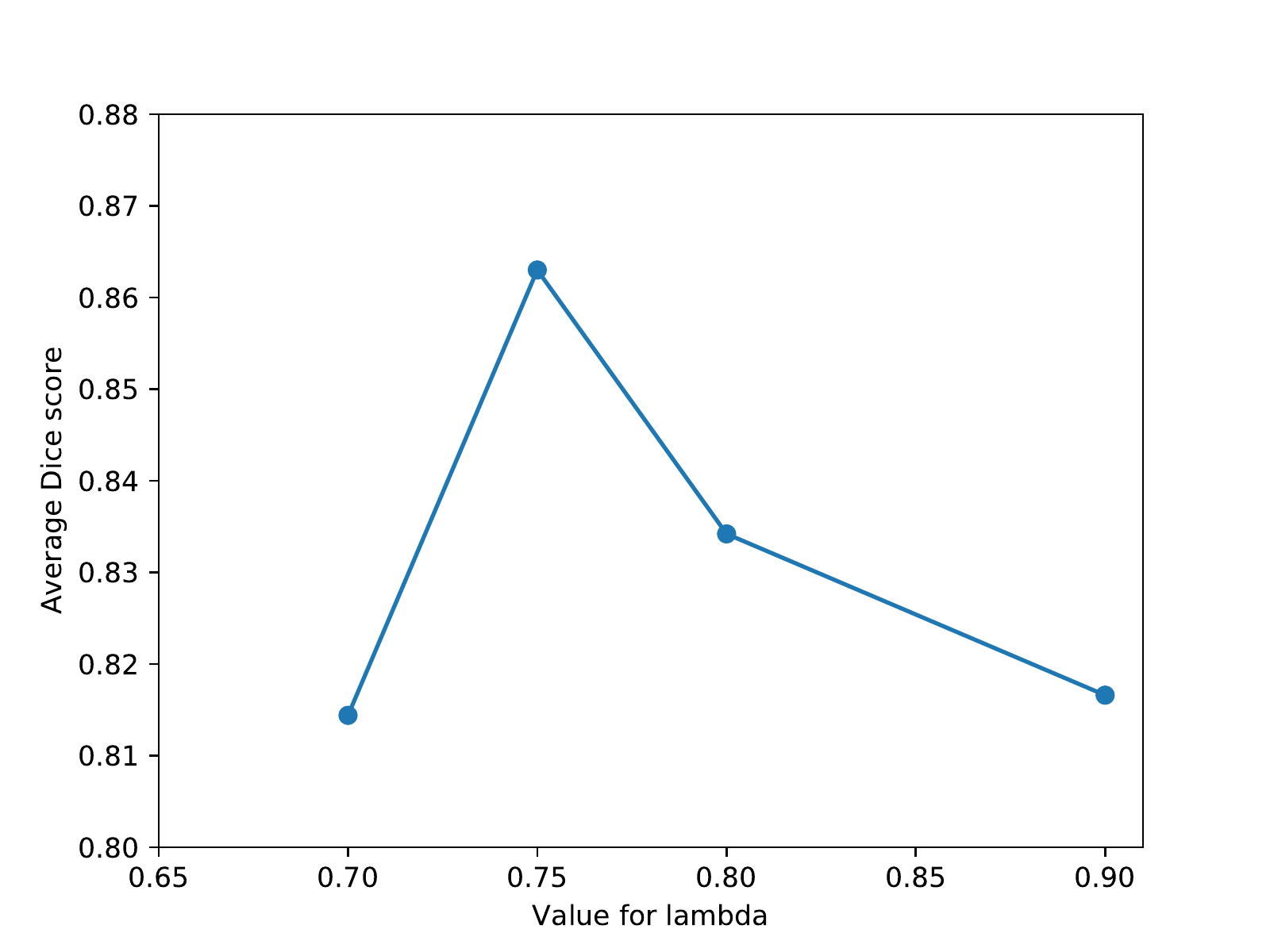}
    \caption{Plot showing optimal lambda($\lambda$) in switching loss for RVSC dataset. The best performance is achieved with $\lambda=0.75$ in the switching loss.}
    \label{fig:radiology_plot}
\end{figure}

Equation \ref{eqn:loss_new} defines the proposed switching loss. It contains a hyperparameter $\lambda$. The hyperparameter $\lambda$ is chosen based on the average imbalance in the dataset per class. For example, if the foreground-class pixel population is 20\% of the total population, then we take $\lambda$ to be 0.8. We demonstrate experimentally that there is an optimal value for $\lambda$ in the proposed switching loss. We train a U-Net model on RVSC dataset using various values of $\lambda$ in the switching loss. Figure \ref{fig:radiology_plot} shows the average Dice score on the validation dataset. The model trained with $\lambda$ equal to 0.75 achieved the best performance, and the performance decreases in both direction of the  $\lambda$ axis.

\subsection{Solving challenges for practical use}

The proposed solution is ready to be integrated in whole slide processing pipelines due to the following features that we have implemented in our software.
 
\subsubsection{Seamless WSI integration via open-source libraries}

We built a software utility that works right out-of-the-box on gigapixel whole slide images that have large file sizes and proprietary formats. It is based on open-source software including OpenSlide library~\cite{goode2013openslide} to read WSIs, PyTorch libraries for programming neural networks, and the python version of the SPAMS toolbox~\cite{mairal2010online} for an improved implementation of SPCN~\cite{vahadane_structure-preserving_2016} as described previously~\cite{ramakrishnan_fast_2019}. Our software utility outputs a CSV file containing the coordinates of the centroids of the detected nuclei as well as an image file with their binary localization masks (which are roughly 70\% shrunk compared to the ground truth, as described previously).

\subsubsection{Run-time on WSIs}
\begin{figure}
    \centering
    \includegraphics[width=12cm]{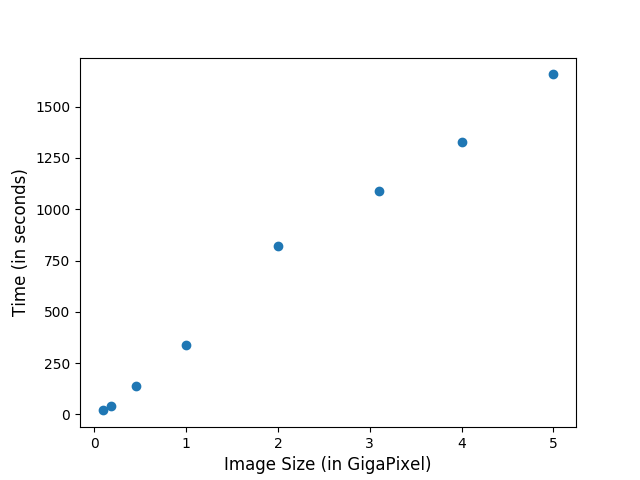}
    \caption{Time consumed versus image size.}
    \label{fig:timewsi}
\end{figure}

We made sure that our software utility was GPU-compatible for inference also, and we tested its speed with different WSI slides. As shown in Figure \ref{fig:timewsi} the processing time scales linearly with the size of the WSI and takes less than six minutes per gigapixel. All experiments were performed on a computer with an 8 core CPU with 64 GB RAM and an Nvidia Titan XP GPU with 12 GB VRAM. 

\subsubsection{Comparison with existing software utilities}
We now compare the features of the proposed solution to existing ready-to-use and publicly available solutions for detecting nuclei, which are Fiji~\cite{fiji_schindelin2012} and CellProfiler~\cite{carpenter2006cellprofiler}. As shown in Table \ref{tab:feature_comp}, these two alternative solutions are general purpose software for building image processing pipelines for biomedical images and have their graphical user interfaces (GUIs). However, for detecting nuclei, these two software can neither open WSIs, nor provide reasonable individual detection of crowded nuclei (see~\cite{kumar_dataset_2017}), nor utilize GPUs for speed-up. Our open-source software utility is not general purpose, but it specializes in directly working on small and large images, including those in proprietary WSI formats, by incorporating our pre-trained neural networks that is rigorously tested for accuracy and speed. The executables and codes of this work are available at \url{https://github.com/MEDAL-IITB/Nuclei_detection}. We have inherited the licenses from the respective open-source libraries used in this project.

\begin{table}
\caption{Software feature comparison with Fiji~\cite{fiji_schindelin2012}, and  CellProfiler~\cite{carpenter2006cellprofiler}.}
    \label{tab:feature_comp}
    \begin{center}
    \begin{tabular}{|c|c|c|c|}
        \hline
        Feature & Fiji & CellProfiler & Ours\\
        \hline
        Opens WSI & \xmark & \xmark & \cmark\\
        directly & & & \\
        \hline
        Detects nuclei & \xmark & \xmark & \cmark\\
         with high & & & \\
         accuracy & & & \\
        \hline
        Uses GPU & \xmark & \xmark & \cmark\\
        acceleration & & & \\
        \hline
        Output format & Mask & Mask & CSV+Mask\\
        \hline
        GUI & \cmark & \cmark & \xmark\\
        \hline
        Can do tasks & \cmark & \cmark & \xmark\\
        other than & & & \\
        nucleus detection & & &\\
        \hline
    \end{tabular}
    \end{center}
\end{table}

\section{Conclusion}
\label{sec:discuss}
We proposed and released a pre-trained nucleus segmentation method that is ready for inferencing right out of the box. We addressed the problem of class-imbalance in nucleus detection by proposing a novel switching loss function to adapt the emphasis given to the foreground or background automatically. We also found that the usage of nuclear stain channel leads to higher detection accuracy than color normalization or raw images. Our method achieved state-of-art results on a multi-organ dataset~\cite{kumar_dataset_2017}, even when the scale of the images was varied. We also achieved state-of-the-art performance on TNBC~\cite{naylor2017nuclei}, and appealing qualitative results on CPCTR~\cite{melamed_cooperative_2004}, CRC Dataset~\cite{sirinukunwattana_locality_2016} and Warwick~\cite{qaiser_her2_2018} datasets without retraining. The proposed method detects a wide range of nuclei, including those of the stroma, epithelium, adipose, and the lymphatic system. It even performed reasonably well on IHC images without having to train on IHC. Additionally, the proposed loss function also showed evidence of being applicable beyond nucleus detection by working well for right ventricle segmentation in MRI. 

We presented a comparison of four popular FCN architectures viz. VGG-UNet, UNet, SegNet, and GCN. The proposed method also uses a computationally inexpensive but effective approach to separate touching nuclei by shrinking the ground truth masks towards the respective centroids of the nuclei. Additionally, we used GPU acceleration for stain separation and color normalization as well, even though these steps do not use deep learning. The entire method has been packaged into a software utility that performs with usable speed on large WSIs, including those saved using proprietary file formats. We hope that the proposed switching loss and nucleus detection software will help further research in computational pathology.

In the future, we will work on adding more features that build on top of the software release. Plans include a graphical user interface to overlay the detection results on WSIs. Nucleus segmentation for morphometrics is also a natural next step. Additionally, we also plan to add modules to train nucleus classification into cell-types such as lymphocytes, epithelial nuclei, stromal nuclei, and mitotic nuclei for H\&E stained images. We invite the research community to collaborate and add to this project and make it a shared public asset.

\section{Disclosures}
Authors have no conflict-of-interest to declare.

\section{Acknowledgements}
Authors thank the Nvidia Corporation for its GPU grant, and Amit More for suggesting improvements to the manuscript.
\bibliography{sample.bib}   
\bibliographystyle{spiejour}   

\listoffigures
\listoftables
\end{spacing}
\end{document}